\documentclass{iau} 
\usepackage{graphicx}
\usepackage{natbib}
\usepackage{hyperref}

\title[Massive stars in metal-poor dwarf galaxies are often extreme rotators] 
{Massive stars in metal-poor dwarf galaxies are often extreme rotators}

\author[Abel Schootemeijer \textit{et al.}]   
{Abel Schootemeijer$^{1}$,
 Danny J. ~Lennon$^{2,3}$,
 Miriam Garcia$^{4}$,
 Norbert Langer$^{1,5}$,
 Ben Hastings$^{1,5}$,
 \and Christoph Sch\"{u}rmann$^{1,5}$}

\affiliation{$^1$Argelander-Institut f\"{u}r Astronomie, Universit\"{a}t Bonn, Auf dem H\"{u}gel 71, 53121 Bonn, Germany
\\[\affilskip]
$^{2}$Instituto de Astrof\'{i}sica de Canarias, E-38200 La Laguna, Tenerife, Spain
\\[\affilskip]
$^3$Departamento de Astrof\'{i}sica, Universidad de La Laguna, E-38205 La Laguna, Tenerife, Spain
\\[\affilskip]
$^4$Centro de Astrobiolog\'{i}a, CSIC-INTA. Crtra. de Torrej\'{o}n a Ajalvir km 4, E-28850 Torrej\'{o}n de Ardoz (Madrid), Spain
\\[\affilskip]
$^5$Max-Planck-Institut f\"{u}r Radioastronomie, Auf dem H\"{u}gel 69, 53121 Bonn, Germany
}

\pubyear{2022}
\volume{361}  
\jname{Massive Stars Near and Far}
\editors{N.~St-Louis, J.~S.~Vink \& J.~Mackey, eds.}
\begin{document}

\maketitle

\begin{abstract}
We probe how common extremely rapid rotation is among massive stars in the early universe by measuring the OBe star fraction in nearby metal-poor dwarf galaxies. We apply a new method that uses broad-band photometry to measure the galaxy-wide OBe star fractions in the Magellanic Clouds and three more distant, more metal-poor dwarf galaxies. We find OBe star fractions of $\sim$20\% in the Large Magallanic Cloud (0.5$Z_\odot$), and $\sim$30\% in the Small Magellanic Cloud (0.2$Z_\odot$) as well as in the so-far unexplored metallicity range $0.1 \lesssim Z/Z_\odot < 0.2$ occupied by the other three dwarf galaxies. Our results imply that extremely rapid rotation is common among massive stars in metal-poor environments such as the early universe.
\end{abstract}

\firstsection 
\section{Introduction}
Massive stars in the early universe are thought to be metal-poor because after the Big Bang, the universe consisted almost exclusively of hydrogen and helium.
Nearby dwarf galaxies are unique laboratories to study the early universe because they are both metal poor \citep{Mateo98} and nearby enough to resolve individual massive stars. The evolution of these massive stars at low metallicity is thought to be strongly impacted by extremely rapid rotation \citep[e.g.,][]{Langer92}.
In line with this is that superluminous supernovae, long-duration gamma-ray bursts, and ultra-luminous X-ray sources \citep[all of which have been linked to rapid rotation - e.g.,][]{Aguilera18, Marchant16} are observed to occur mainly in low-metallicity dwarf galaxies \citep{Lunnan14, Kaaret17}.
Diagnostics for extremely rapid rotation in stars are the presence of H$\alpha$ emission and excess flux at long wavelengths, typically thought to originate from a decretion disk \citep[see, e.g.,][]{Struve31, Poeckert76, Vink09}. Stars that display these features are classified as OBe stars. Here we investigate the prevalence of such rapidly rotating OBe stars in nearby metal-poor dwarf galaxies.

\begin{figure}[ht]
\begin{center}
 \includegraphics[width=0.9\linewidth]{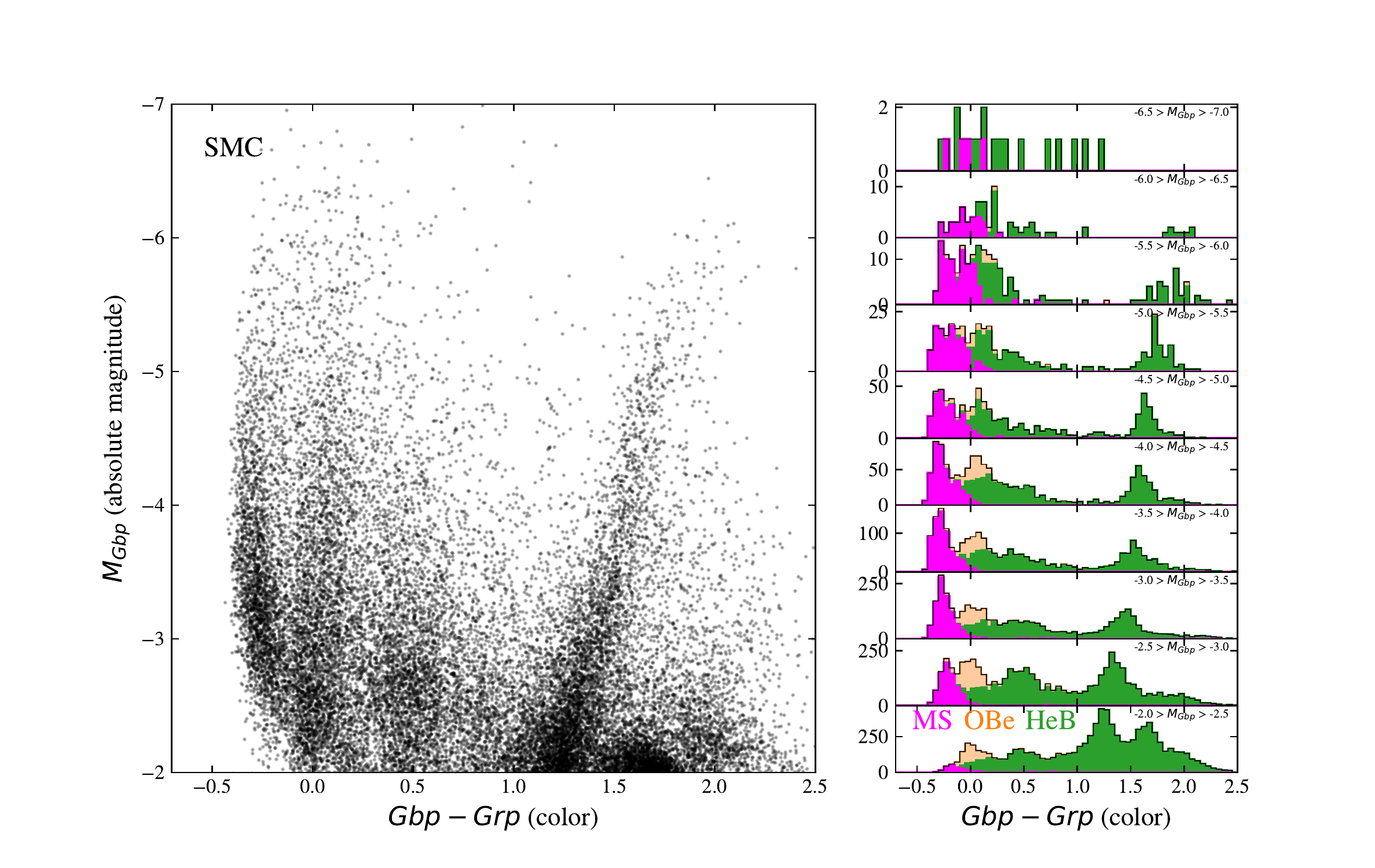} 
 \caption{\textit{Left:} color-magnitude diagram of bright sources in the Small Magellanic Cloud.
\textit{Right:} stacked color histograms showing the number of sources within the main sequence (MS, magenta), OBe (orange), and helium-burning (HeB, green) groups, as identified by the color-color cuts displayed in Fig.\,\ref{fig:coco_smc}.}

   \label{fig:cmd_smc}
\end{center}
\end{figure}

\section{Methods}
We employ archival broad-band photometry to separate OBe stars, main sequence (MS) stars without disks, and cooler helium-burning (HeB) stars. We use colors at short wavelengths to identify hot stars and colors at longer wavelengths to infer the presence of a disk. Below we describe the data sets used in this work.

\subsection{Magallanic Cloud data sets}
For the Small and Large Magellanic Cloud (SMC and LMC), we cross-correlate\footnote{Using \url{http://cdsxmatch.u-strasbg.fr/} 
} the data from GAIA EDR3 \citep{Gaia21}, with classical $UB$ photometry \citep{Zaritsky02, Zaritsky04} and infrared Spitzer-SAGE photometry \citep{Meixner06}, adopting a 1" radius. 

\subsection{Sextans\,A, Holmberg\,I, and Holmberg\,II data sets}
Sextans\,A, Holmberg\,I, and Holmberg\,II are about twenty to a hundred times further away than the Magellanic Clouds. For that reason, only data from the Hubble Space Telescope (HST) is deep enough for our method while at the same time having a high enough angular resolution.
For Sextans\,A, we cross-correlate\footnote{Using the Aladin sky atlas 
}
$F336W$ and $F439W$ data from \cite{Bianchi12} with $F555W$ and $F814W$ data from \cite{Holtzman06}. 
Before cross-correlating, we shift the data from \cite{Bianchi12} by 0.6" in right ascension and 1" in declination. The cross-correlation radius is 0.35".
For Holmberg\,I and Holmberg\,II we use the photometric data from the Legacy ExtraGalactic UV Survey \citep[LEGUS;][]{Sabbi18}.

\begin{figure}[ht]
\begin{center}
 \includegraphics[width=\linewidth]{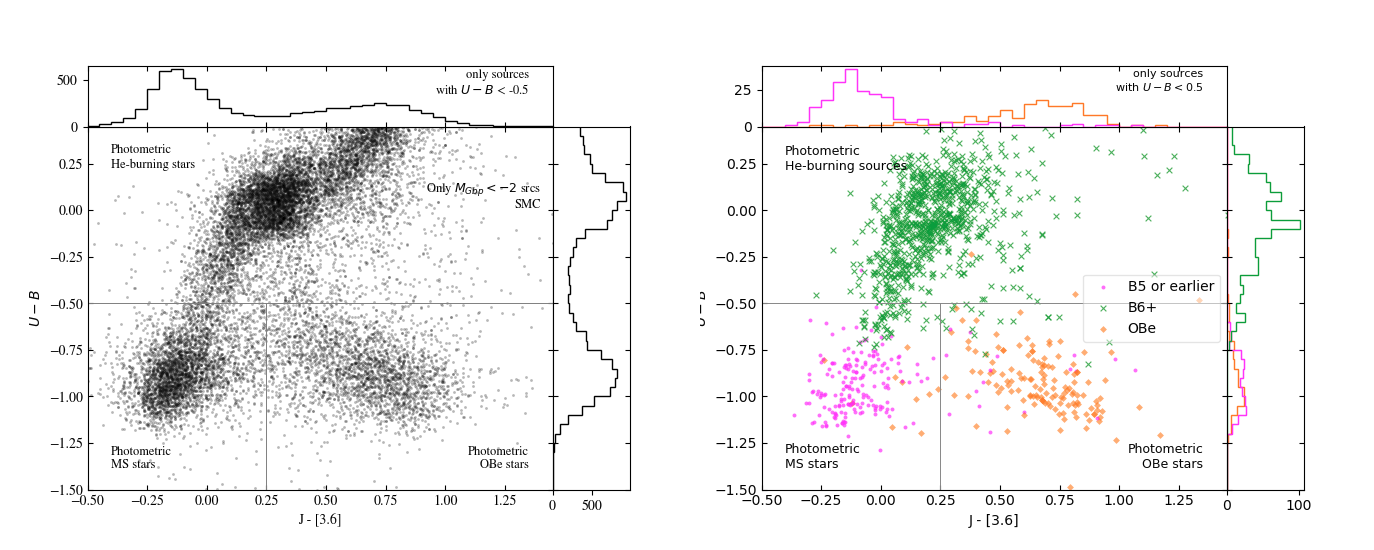} 
 \caption{Color-color diagrams with histograms of both colors shown on the top and right side. \textit{Left:} all sources brighter than $M_{Gbp} < -2$ in our data set. \textit{Right:} same, but showing only sources for which spectral types have been determined (see text for more details).}
   \label{fig:coco_smc}
\end{center}
\end{figure}

\section{Results}
\subsection{The Magellanic Clouds}
The color-magnitude diagram (CMD) of the SMC (Fig.\,\ref{fig:cmd_smc}) shows several features. Here we focus on the two bluest bands, at $Gbp-Grp \approx -0.25$ and $Gbp-Grp \approx 0$. To investigate the nature of the sources in these bands, we plot the sources shown in our CMD also in a color-color diagram (Fig.\,\ref{fig:coco_smc}, left). In this diagram, three groups form. 
At the right side of Fig.\,\ref{fig:coco_smc}, we investigate their nature by plotting only sources for which spectral types have been determined \citep{Martayan07, Hunter08b, Bonanos10, Dufton19}. 
Sources with $U-B > -0.5$ typically have spectral types of B6 and later, which is indicative of an effective temperature of $T_\mathrm{eff} < 15$\,kK \citep{Schootemeijer21}. Such cool stars should be HeB stars, because the MS in the SMC does not extend down to temperatures this low \citep{Schootemeijer19}.
The sources with $U-B < -0.5$ and $J-[3.6] < 0.25$ tend to have spectral types of B5 and earlier, which implies that they are MS stars. The group of sources to the right of them consists almost exclusively of OBe stars \citep[at  a similar infrared color as found by][]{Bonanos10}. Given how well these three groups of stars are separated in the color-color diagram, we conclude that this diagram can be used efficiently to tell apart MS, OBe, and HeB sources.

Turning back to Fig.\,\ref{fig:cmd_smc}, we see in the right panels that the bluest band with $Gbp-Grp \approx -0.25$ exclusively consists of MS stars without decretion disks. The band parallel to the MS band consists of a combination of HeB stars, and, perhaps more surprisingly, a significant number of OBe stars. We will quantify the fraction of OBe stars later on.
The relatively high number of bright HeB stars -- compared to MS stars -- is explained by HeB stars being relatively bright in the visual.
In the LMC, we use the same method to identify HeB, MS, and OBe stars, except that we place the threshold for HeB stars at $U-B > -0.3$ to account for redder intrinsic colors and higher extinction.

\begin{figure}[ht]
\begin{center}
 \includegraphics[width=0.635\linewidth]{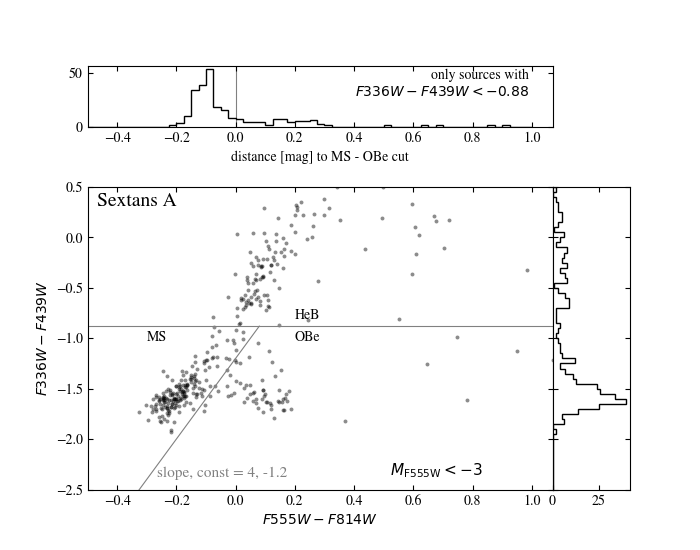} 
 \caption{Color-color diagram of bright sources ($M_\mathrm{F555W} < -3$) in our Sextans\,A data set. On the right we show a histogram of their $F336W - F439W$ colors. At the top we show a histogram of the shortest distance to the main sequence (MS) - OBe cut of the sources that have $F336W - F439W < -0.88$. Sources with $F336W - F439W > -0.88$ are considered to be helium-burning (HeB) sources.}
   \label{fig:coco_sexa}
\end{center}
\end{figure}

\begin{figure}[ht]
\begin{center}
 \includegraphics[width=0.635\linewidth]{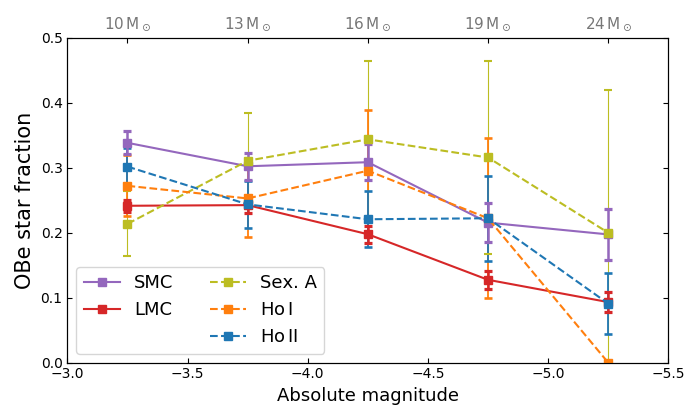} 
 \caption{OBe star fractions as a function of absolute magnitude, measured in intervals of half a magnitude. We show these for the Small Magellanic Cloud (SMC), Large Magellanic Cloud (LMC), Sextans\,A (Sex.\,A), Holmberg\,I (Ho\,I), and Holmberg\,II (Ho\,II). At the top we show, as an indication, average evolutionary masses in the same absolute magnitude intervals (see text).}
   \label{fig:fobe}
\end{center}
\end{figure}

\begin{figure}[ht]
\begin{center}
 \includegraphics[width=0.61\linewidth]{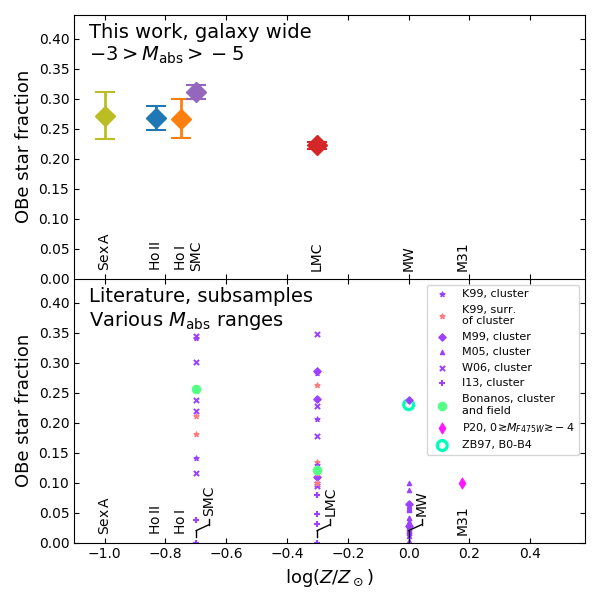} 
 \caption{Metallicity trends of the OBe star fraction. 
 \textit{Top:} this work, in the absolute magnitude range $-3 > M_\mathrm{abs} > -5$. \textit{Bottom:} literature OBe star fractions within stellar sub-samples of different galaxies, measured in various $M_\mathrm{abs}$ ranges. We also show Milky Way (MW) and M31 data.  Included studies are K99 \citep{Keller99}, M99 \citep{Maeder99}, M05 \citep{McSwain05}, W06 \citep{Wisniewski06}, I13 \citep{Iqbal13}, Bonanos \citep{Bonanos09, Bonanos10}, P20 \citep{Peters20}, and ZB97 \citep{Zorec97}. The Sex\,A metallicity is based on \cite{Kaufer04}; for Ho\,I and Ho\,II and Ho\,II it is based on \cite{Croxall09}.}
   \label{fig:ztrends}
\end{center}
\end{figure}

\subsection{Sextans\,A, Holmberg\,I, and Holmberg\,II}
To identify MS, OBe, and HeB stars in Sextans\,A using HST data, we employ the color-color diagram shown in Fig.\,\ref{fig:coco_sexa}. As in Fig.\,\ref{fig:coco_smc}, we see that three groups form, which we interpret as HeB, MS, and OBe stars. We have verified the OBe nature of the stars in the `OBe' region of the color-color diagram in two ways: i) in the SMC, stars with known OBe spectral types form a separate group in a color-color diagram with colors similar to Fig.\,\ref{fig:coco_sexa} ($Gbp-Grp$ versus $U-B$), and, ii), the sources in Sextans\,A identified by us as OBe stars almost always have H$\alpha$ in emission in according to narrow-band $F656N$ photometry of \cite{Dohm-Palmer02}.

For Holmberg\,I and Holmberg\,II, we apply the same strategy as for Sextans\,A, except that we use from the LEGUS data a $V-I$ color instead of $F555W - F814W$ and $U-B$ instead of $F336W - F439W$. 

\subsection{OBe star fractions as a function of absolute magnitude}

In intervals of half a magnitude, based on our color-color diagrams we calculate the fraction of main sequence stars that are OBe stars: $f_\mathrm{OBe} = N_\mathrm{OBe} / (N_\mathrm{MS} + N_\mathrm{OBe})$.
We do this for the SMC, LMC, Sextans\,A, Holmberg\,I, and Holmberg\,II. 
The result is shown in Fig.\,\ref{fig:fobe}. We find high OBe star fractions of $f_\mathrm{OBe} \approx 0.2 - 0.3$, except at the bright end, where the OBe star fractions tend to decrease. This decrease is the strongest in the LMC, which could be related to stellar winds being relatively strong there (as the LMC is the highest-metallicity galaxy in our sample).
As an indication, 
we write the average evolutionary masses \citep{Schootemeijer21} that have been determined for stars with spectral types of B5 and earlier in the SMC at the top of Fig.\,\ref{fig:fobe}. Typical average masses in the considered absolute magnitude range vary from 10\,M$_\odot$ to 25\,M$_\odot$. 

\subsection{OBe star fractions as a function of metallicity}
To investigate metallicity trends of the OBe star fraction, we consider one absolute magnitude range: $-3 > M_\mathrm{abs} > -5$. The top panel of Fig.\,\ref{fig:ztrends} shows our result. At $ 0.1 \leq Z / Z_\odot \leq 0.2$ (i.e., in Sex\,A, Ho\,I, Ho\,II, and the SMC) as much as $\sim$30\% of the hydrogen-burning sources is an OBe star. In the LMC, we find a value of $\sim$20\%.

In the LMC and the SMC, we can compare our results with studies from the literature. 
The OBe star fraction has been studied before in different environments, albeit never in a galaxy-wide sample as we did here. Instead, typically a sub-sample of stars in a star cluster has been targeted. As the bottom panel of Fig.\,\ref{fig:ztrends} shows, the result then depends on the cluster that is picked (e.g., because of cluster age), and also on which of its stars are included in the measurement. In the LMC and SMC, our OBe star fractions lie at the high end of the literature values.

\section{Concluding remarks}
We have devised a method that uses broad-band photometry to tell apart HeB, MS, and OBe stars. With data from the archive, we have determined galaxy-wide OBe star fractions. 
These include the first OBe star fractions at sub-SMC metallicty in Sextans\,A, Holmberg\,I, and Holmberg\,II. The latter two belong to the M81 group, meaning that OBe star fraction estimates have been extended beyond the Local Group. 
We found that in the range $0.1-0.2$Z$_\odot$, as much as thirty per cent of massive stars is an extremely rapid rotator. This result implies that a large fraction of massive stars in the early universe could be rapid rotators, and it sheds light on the preference for metal-poor dwarf galaxies that long-duration gamma-ray bursts, superluminous supernovae, and ultraluminous X-ray sources have. 


\bibliographystyle{apj}
\bibliography{./bib}

\begin{discussion}

\discuss{(Couldn't hear a name sorry)}{
In Sextans\,A and the like where you do not have spectral types, how well can you separate early and late B stars?
}

\discuss{Schootemeijer}{
I agree that there is uncertainty in the exact location of the border between these stars. What I can say is that most of the O and early-B stars reside in a well-defined group, so the exact value that is chosen for the border should not affect our derived OBe star fraction too much.
}

\end{discussion}

\end{document}